\documentclass[11pt]{article}
\usepackage{times}
\usepackage{sidecap}
\usepackage{geometry}
\usepackage{floatrow}
\usepackage{float}
\usepackage[breaklinks=true,colorlinks=true,citecolor=blue,linkcolor=blue,urlcolor=blue]{hyperref}
\usepackage{graphics,graphicx}
\usepackage{subcaption}
\usepackage{epsf,wrapfig}
\usepackage{tcolorbox}

\usepackage{natbib,bm}

\usepackage{paralist}
\let\olditem\item
\renewenvironment{thebibliography}[1]{%
  \section*{\refname}
  \let\par\relax
  \renewcommand{\item}[1][]{\olditem[\textbullet]}%
  \inparaenum}{\endinparaenum}

\newcounter{aff}

\DeclareGraphicsExtensions{.jpg,.pdf,.png,.eps}

\usepackage{caption}
\usepackage{xcolor}

\usepackage{tcolorbox}

\bibpunct{(}{)}{;}{a}{}{,}

\usepackage{soul}

\definecolor{darkcyan}{rgb}{0.0, 0.0, 0.7}

\definecolor{boxcolor}{RGB}{193, 229, 255}
\definecolor{figboxcolor}{RGB}{193, 229, 255}
\definecolor{boxcolorgray}{RGB}{227, 227, 227}
\definecolor{darkcyan}{rgb}{0.0, 0.0, 0.7}

\sethlcolor{boxcolorgray}

\def\gtrsim{\lower 2pt \hbox{$\, \buildrel {\scriptstyle >}\over
{\scriptstyle \sim}\,$}}
\def\lesssim{\lower 2pt \hbox{$\, \buildrel {\scriptstyle <}\over
{\scriptstyle \sim}\,$}}

\def\approxlt{\lower.2em\hbox{$\buildrel < \over \sim$}}
\def\approxgt{\lower.2em\hbox{$\buildrel > \over \sim$}}

\def\be{\begin{equation}}
\def\ee{\end{equation}}
\def\bea{\begin{eqnarray*}}
\def\eea{\end{eqnarray*}}

\def\jref@jnl#1{{\rm#1}}
\def\aj{\jref@jnl{AJ}}			
\def\araa{\jref@jnl{ARA\&A}}		
\def\apj{\jref@jnl{ApJ}}			
\def\apjl{\jref@jnl{ApJ}}		
\def\apjs{\jref@jnl{ApJS}}		
\def\ao{\jref@jnl{Appl.~Opt.}}		
\def\apss{\jref@jnl{Ap\&SS}}		
\def\aap{\jref@jnl{A\&A}}		
\def\aapr{\jref@jnl{A\&A~Rev.}}		
\def\aaps{\jref@jnl{A\&AS}}		
\def\azh{\jref@jnl{AZh}}			
\def\baas{\jref@jnl{BAAS}}		
\def\jrasc{\jref@jnl{JRASC}}		
\def\memras{\jref@jnl{MmRAS}}		
\def\mnras{\jref@jnl{MNRAS}}
\def\nar{\jref@jnl{Nat.~Astr.~Rev.}}
\def\pra{\jref@jnl{Phys.~Rev.~A}}	
\def\prb{\jref@jnl{Phys.~Rev.~B}}	
\def\prc{\jref@jnl{Phys.~Rev.~C}}	
\def\prd{\jref@jnl{Phys.~Rev.~D}}	
\def\pre{\jref@jnl{Phys.~Rev.~E}}	
\def\prl{\jref@jnl{Phys.~Rev.~Lett.}}	
\def\pasp{\jref@jnl{PASP}}		
\def\pasj{\jref@jnl{PASJ}}		
\def\qjras{\jref@jnl{QJRAS}}		
\def\skytel{\jref@jnl{S\&T}}		
\def\solphys{\jref@jnl{Sol.~Phys.}}	
\def\sovast{\jref@jnl{Soviet~Ast.}}	
\def\ssr{\jref@jnl{Space~Sci.~Rev.}}	
\def\zap{\jref@jnl{ZAp}}			
\def\nat{\jref@jnl{Nature}}		
\def\iaucirc{\jref@jnl{IAU~Circ.}}
\def\aplett{\jref@jnl{Astrophys.~Lett.}}
\def\apspr{\jref@jnl{Astrophys.~Space~Phys.~Res.}}
\def\bain{\jref@jnl{Bull.~Astron.~Inst.~Netherlands}}
\def\fcp{\jref@jnl{Fund.~Cosmic~Phys.}}
\def\gca{\jref@jnl{Geochim.~Cosmochim.~Acta}}
\def\grl{\jref@jnl{Geophys.~Res.~Lett.}}
\def\jcp{\jref@jnl{J.~Chem.~Phys.}}	
\def\jgr{\jref@jnl{J.~Geophys.~Res.}}	
\def\jqsrt{\jref@jnl{J.~Quant.~Spec.~Radiat.~Transf.}}
\def\memsai{\jref@jnl{Mem.~Soc.~Astron.~Italiana}}
\def\nphysa{\jref@jnl{Nucl.~Phys.~A}}
\def\physrep{\jref@jnl{Phys.~Rep.}}
\def\physscr{\jref@jnl{Phys.~Scr}}
\def\planss{\jref@jnl{Planet.~Space~Sci.}}
\def\procspie{\jref@jnl{Proc.~SPIE}}

\def\ion#1#2{#1$\;${\small\rm\@Roman{#2}}\relax}

\newcount\lecurrentfam
\def\LaTeX{\lecurrentfam=\the\fam \leavevmode L\raise.42ex
\hbox{$\fam\lecurrentfam\scriptstyle\kern-.3em A$}\kern-.15em\TeX}
\def\sizrpt{
(\fontname\the\font): em=\the\fontdimen6\font, ex=\the\fontdimen5\font
\typeout{
(\fontname\the\font): em=\the\fontdimen6\font, ex=\the\fontdimen5\font
}}
\definecolor{dark}{rgb}{ 0.34, 0.09, 0.49}

\usepackage{fancyhdr,url}
\usepackage{transparent}
\usepackage{tablefootnote}

\usepackage{mathpazo}

\begin{document}

\noindent
\begin{raggedright}
{\Huge\bfseries Resolving the AGN Feeding and Feedback Cycle with ALMA2040\\[0.5em]
\Large Kinematics and multi-line diagnostics of the cold molecular ISM}\end{raggedright}\\

\noindent

Ilaria Ruffa$^{1,2}$, Francesco Salvestrini$^{3,4}$, Almudena Alonso Herrero$^{5}$, Anelise Audibert$^{6,7}$, Elena Bertola$^1$, Viviana Casasola$^{2}$, Roberto Decarli$^{8}$, Daniel Espada$^{9,10}$, Umberto Maio$^{3,4}$, Dieu D. Nguyen$^{11}$, Mamta Pandey$-$Pommier$^{12}$, Miguel Pereira-Santaella$^{13}$, Francesca Pozzi$^{14,8}$, Jahang Prathap$^{15,16}$, Sabine Thater$^{17}$, Maria Vittoria Zanchettin$^{18}$

\vspace{0.5em}
{
\footnotesize{
\noindent $^1$ INAF - Osservatorio Astrofisico di Arcetri, Largo E. Fermi 5, 50125 Firenze, Italy; 
$^2$ INAF–Istituto di Radioastronomia, Via Gobetti 101, I-40129 Bologna, Italy; 
$^3$ INAF, Osservatorio Astronomico di Trieste, via Tiepolo 11, I-34131, Trieste, Italy; 
$^4$ IFPU - Institute for Fundamental Physics of the Universe, via Beirut 2, I-34151 Trieste, Italy; 
$^5$ Centro de Astrobiología (CAB), CSIC–INTA, Camino Bajo del Castillo s/n, E–28692 Villanueva de la Ca\~nada, Madrid, Spain; 
$^6$ Instituto de Astrofísica de Canarias, Calle Vía Láctea, s/n, E-38205, La Laguna, Tenerife, Spain; 
$^7$ Departamento de Astrofísica, Universidad de La Laguna, E-38206, La Laguna, Tenerife, Spain; 
$^8$ INAF– Osservatorio di Astrofisica e Scienza dello Spazio di Bologna, Via Gobetti 93/3, I-40129 Bologna, Italy; 
$^9$ Dpto. de F\'isica Te\'orica y del Cosmos, Campus de Fuentenueva, Universidad de Granada, E-18071, Granada, Spain; 
$^{10}$ Instituto Carlos I de F\'isica Te\'orica y Computacional, Universidad de Granada, E-18071, Granada, Spain; 
$^{11}$ Department of Astronomy, University of Michigan, 1085 South University Avenue, Ann Arbor, MI 48109, USA; $^{12}$ Pole Scientific, University Catholic of Lyon- University of Lyon, 10 place des Archives 69002, Lyon, France; 
$^{13}$ Instituto de F\'isica Fundamental, CSIC, Calle Serrano 123, 28006 Madrid, Spain
$^{14}$ Dpto. di Fisica e Astronomia, Alma Mater Studiorum Univ. di Bologna, Via P. Gobetti 93/2, I-40129 Bologna, Italy; 
$^{15}$ School of Mathematical and Physical Sciences, Macquarie University, Sydney, NSW 2109, Australia; 
$^{16}$ Macquarie University Astrophysics and Space Technologies Research Centre, Sydney, NSW 2109, Australia; $^{17}$Department of Astrophysics, University of Vienna, Türkenschanzstraße 17, 1180 Vienna, Austria; 
$^{18}$ INAF -- Istituto di Astrofisica e Planetologia Spaziali, Via Fosso del Cavaliere 100, 00133 Rome, Italy
}
}

\begin{tcolorbox}[boxsep=0.0pt,left=0pt,right=0pt,top=1pt,bottom=0.0pt,colframe=boxcolor,colback=boxcolor]
\justify
\textbf{Abstract:} 
Understanding how supermassive black holes (SMBHs) grow and regulate the growth of their host galaxies requires tracing the cold molecular interstellar medium (ISM) throughout the full AGN feeding--feedback cycle. Molecular gas constitutes both the primary fuel for star formation and black-hole accretion and one of the ISM phases most directly affected by AGN-driven radiation, winds, and radio jets. Over the past decade, ALMA has resolved circumnuclear molecular discs, rings, spirals, tori, and jet-driven outflows on scales of tens of parsecs in nearby AGN. These studies have demonstrated that combining spatially resolved gas kinematics with multi-line diagnostics is essential for disentangling the physical processes underlying the observed gas conditions. However, our current picture remains largely based on a limited number of exceptionally nearby systems. Most local AGN have instead been observed with ALMA in a single transition at moderate or low angular resolution, preventing a systematic characterisation of both the dynamics and physical state of the gas. Moreover, existing studies have largely overlooked kinetic-mode AGN hosted by massive early-type galaxies, despite their central importance for understanding maintenance-mode feedback. In this White Paper, we argue that ALMA2040 should transform these detailed case studies into a statistical framework. A $\sim10\times$ improvement in line sensitivity relative to the current ALMA, together with fast mosaicking, ultra-wide instantaneous bandwidth, and longer baselines (resolving up to $\sim9$ mas at 230 GHz), will enable homogeneous, multi-line mapping of the molecular ISM on giant molecular clouds scales across statistically significant samples of both radiative- and kinetic-mode AGN out to $z\lesssim0.1$. 
\end{tcolorbox}

\newpage{}

\vspace{-0.7cm}
\section{The cold molecular ISM as the missing link between SMBH feeding and feedback}

\begin{tcolorbox}
Understanding how supermassive black holes (SMBHs) evolve and regulate the evolution of their host galaxies requires spatially-resolved analyses of both the physics and the kinematics of the cold ISM across the full AGN duty cycle. While ALMA has demonstrated the power of this approach through a handful of nearby case studies, ALMA2040 will transform it into a statistical tool applicable to the full diversity of AGN populations.
\end{tcolorbox}

\subsection{Why cold molecular gas?}

Understanding SMBH-host galaxy co-evolution requires an accurate physical description of two closely interconnected processes: the transport of gas from tens of kpc down to the sphere of influence of the SMBH, and the coupling of the energy and momentum released by accretion with the surrounding interstellar medium (ISM). These two processes, commonly referred to as active galactic nuclei (AGN) feeding and feedback, are not independent phases of galaxy evolution, but they are instead connected processes within the same baryon cycle \citep[e.g.][]{Kormendy13,Harrison17,Veilleux20}. The gaseous ISM must indeed lose angular momentum over many orders of magnitude in radius before it can reach the central SMBH, while the resulting AGN activity can heat, compress, accelerate, or chemically alter the galactic gas reservoirs, thus affecting the subsequent host galaxy evolution 
\citep[e.g.][]{Storchi19}.

The cold molecular ISM is central to this cycle. It constitutes the immediate reservoir for both star formation and SMBH fuelling, and it is also one of the phases most directly affected by AGN-driven radiation, winds, and radio jets \citep[e.g.][]{Feruglio10,Cicone14,Fiore17,Oosterloo17,Lutz20,Ruffa22,Audibert23,Ruffa24}. Unlike ionised or hot gas, which is typically strongly affected by AGN feedback, molecular gas can simultaneously encode the pre-feedback fuel supply, the onset of inflow, and the response of the dense and cold ISM to mechanical or radiative energy injection. In this sense, the molecular gas phase is not simply a tracer of AGN activity: it is the physical medium through which the feeding--feedback loop is established \citep[e.g.][]{GarciaBurillo24}.

Spatially-resolved observations of molecular gas are therefore essential. Carbon monoxide (CO) traces the bulk of the molecular reservoir \citep[e.g.][]{Bolatto13}, allowing the distribution and kinematics of the cold ISM to be mapped on both circumnuclear and galactic scales. Tracers such as HCN, HCO$^+$, HNC, and CS probe molecular gas at higher densities (i.e. $n_{\rm crit}>10^{4}$\,cm$^{-3}$) and most directly exposed to nuclear activity, while molecules such as SiO and HNCO can provide diagnostics of shocked and mechanically processed gas associated with expanding radio jets or radiation-driven outflows  \citep[e.g.][]{Gao04,Meijerink07,Meijerink08,Aalto12,Viti14,Topal16,Jimenez17,Kam18,Jimenez-Donaire19,Holdship21,Rybak26}. The combined and spatially-resolved study of the molecular gas kinematics and multi-line diagnostics thus offers a powerful route to determine not only the role of cold gas in powering the nuclear activity, but also what physical processes are acting upon it.

\begin{figure}[]
\centering
\vspace{0.1cm}
\includegraphics[width=1.0\textwidth]{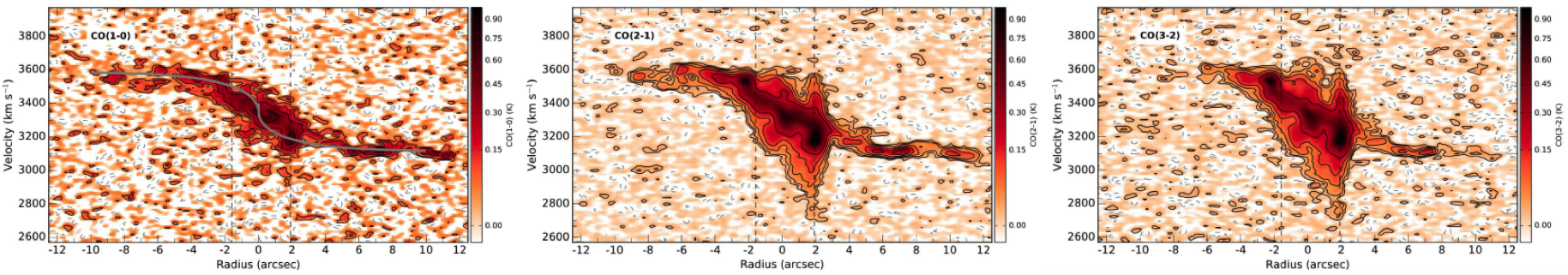}
\vspace{-0.7cm}
\caption{\it From left to right, position--velocity diagrams along the radio-jet axis for CO(1--0), CO(2--1), and CO(3--2), respectively, averaged over 1.5 arcsec perpendicular to the jet in IC~5063. Dashed vertical lines mark the approximate extent of the radio continuum structure, while the grey curve in the CO(1--0) panel indicates the expected locus of regularly rotating disk gas. Adapted from \cite{Oosterloo17} }\label{fig:PV_CO_transitions}
\end{figure}

\subsection{What ALMA has taught us}\label{sec:ALMA_state}
Over the past decade, ALMA has transformed the study of cold gas in AGN. Its angular resolution, sensitivity, and imaging fidelity have made it possible to resolve molecular gas distributions on scales of tens of parsecs in nearby active galaxies, revealing a wide diversity of circumnuclear structures (rotating disks, rings, bars, nuclear spirals, clumpy streams, warped disks, and compact tori; \citealp[e.g.][]{Combes13,Garcia16,AlonsoHerrero18,Imanishi18,Ruffa19a,Ruffa19b, Kameno20,GarciaBurillo21,Ruffa24}), and providing key evidence of the relative impact of AGN and SF-related processes as major mechanisms perturbing the cold gas \citep[e.g.][]{Audibert21, GarciaBurillo24}.

A particularly important lesson from ALMA is that AGN feedback can couple efficiently to the molecular gas phase. AGN-driven molecular outflows have indeed been detected in a variety of local systems, from luminous quasars \citep[e.g.][]{RamosAlmeida22,Audibert23} and ultra-luminous infrared galaxies (ULIRGs; \citealp[e.g.][]{Pereira18, Lamperti22, Holden24,Ruffa26}), to Seyferts and radio galaxies \citep[e.g.][]{Feruglio10,Cicone14,Garcia14,Morganti15,Audibert19,Veilleux20,Ruffa22}. In some of these systems, the molecular gas appears to be entrained or accelerated by radiative AGN feedback in the form of wide-angle winds; in others, the dominant driver appears to be the interaction between radio plasma (i.e. the radio jets) and a clumpy multiphase ISM. For instance, the case of the nearby Seyfert IC~5063 has become a benchmark example of a radio jet driving a fast molecular outflow and strongly modifying the excitation conditions of the impacted gas \citep[][see Fig.~\ref{fig:PV_CO_transitions}]{Morganti15,Dasyra16,Oosterloo17}. High-resolution observations of other nearby Seyferts (e.g.\,NGC~1068) have also revealed complex circumnuclear molecular gas dynamics in which inflow, outflow, shocks, streaming motions, radiative feedback and turbulence in general are intertwined on scales of only a few tens of parsecs \citep[e.g.][see also White Paper of this series by Pereira-Santaella et al.]{Garcia14,Imanishi18,Audibert19,GarciaBurillo21}. 

ALMA has also opened a new window on the cold gas content of early-type galaxies (ETGs) hosting radio AGN (see also Section~\ref{sec:radio_AGN}). Once thought to be largely devoid of cold gas, local ETGs are now known to host significant molecular reservoirs in a substantial fraction of cases \citep[$\sim25\%$; see e.g.][]{Davis19}. In radio galaxies, ALMA observations have revealed circumnuclear molecular disks, rings, unsettled structures, and jet-affected gas, demonstrating that the cold ISM is directly involved in the life cycle of kinetic-mode AGN \citep[][]{Ruffa19b,Ruffa19a,Ruffa20,Ruffa22,North21,Audibert22,Tadhunter24,Ruffa24,Oosterloo25}. .

Despite these advances, the most detailed studies remain limited to a small number of very nearby ($z\lesssim0.02$), bright and well-resolved systems.\,Therefore, we are still missing a comprehensive view of how the feeding--feedback cycle depends on AGN luminosity, accretion state, radio power, host morphology, environment or evolutionary stage. The next step is to shed light on this matter by studying well-defined AGN populations in the local Universe, providing also a physically-calibrated basis for complementary studies at higher redshift.

\subsection{From fuel to feedback: a physical census of the molecular gas cycle}\label{sec:fuel_feedback}

The central science goal for ALMA2040 in this field is to trace the same molecular gas from its transport towards the SMBH through its subsequent response to the energy and momentum released by accretion, in large samples of both radiative- and kinetic-mode AGN. This requires two complementary observational approaches:\,spatially-resolved gas kinematics and multi-line diagnostics. The former provides direct access to the dynamical state of the cold ISM: velocity fields, position--velocity diagrams, line profiles, and full three-dimensional modelling of molecular data cubes can indeed distinguish ordered rotation from non-circular motions \citep[e.g.\,inflows, outflows, streaming along bars, turbulence; see e.g.][]{Davis17,Ruffa19a,GarciaBurillo21,Ruffa22}. This information is crucial for determining whether molecular gas can lose enough angular momentum to fuel the SMBH, whether it is being accelerated or stirred by the AGN, and whether the affected material escapes the galaxy or remains gravitationally bound.

\begin{wrapfigure}{r}{0.42\textwidth}
    \centering
    \includegraphics[width=\linewidth]{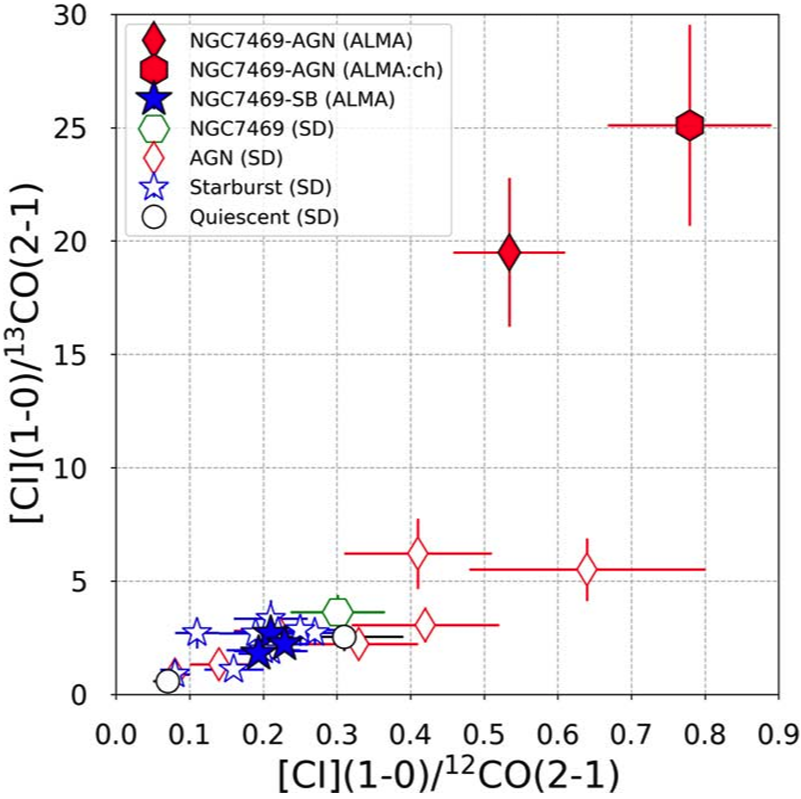}
    \caption{\it [C\,{\sc i}](1--0)/$^{13}$CO(2--1) versus [C\,{\sc i}](1--0)/$^{12}$CO(2--1) line ratios for NGC~7469 and comparison samples \citep[adapted from][]{Izumi20}. Filled symbols show ALMA measurements of the AGN and of the circumnuclear starburst ring (SB) of NGC~7469, with ratios derived from integrated fluxes or, for the AGN, from channel maps (ch).\,Open symbols show single-dish (SD) measurements of NGC~7469 and of AGN, starburst and quiescent galaxies.} 
    \label{fig:izumi20}
\end{wrapfigure}

Kinematics alone, however, cannot reveal how feedback changes the physical state of the gas. AGN radiation (hard X-rays, intense UV fields, enhanced ionisation) can penetrate deep into molecular clouds, altering their temperature, excitation and chemistry, and potentially their ability to form stars. Radio jets can produce similar effects. Crucially, a jet or wind may barely affect the gas kinematics while strongly altering its density and temperature \citep[e.g.][]{Ruffa22}, whereas disturbed kinematics alone cannot reveal whether the gas has been shocked, compressed or thermally processed \citep[e.g.][]{Oosterloo17}. Spatially-resolved multi-line observations are therefore essential to quantify how AGN feedback couples to the cold ISM and distinguish radiative or mechanical heating from other processes \citep[e.g.][see also White Papers in this series by Bouvier et al. and den Brok et al.]{Meijerink05,Viti14,Topal16,Holdship21,Audibert25}.

Within this context, CO spectral-line energy distributions (CO-SLEDs) provide one route to separating the bulk cold reservoir, primarily traced by low-$J$ transitions, from warmer and more highly excited molecular components, typically traced by mid- and high-$J$ CO emission  \citep[e.g.][]{Carilli13,Pozzi17,Esposito22,Esposito24}. Yet high excitation alone is not an unambiguous signature of AGN feedback, as compact star formation, shocks and turbulent heating can produce similar CO ladders \citep[e.g.][]{Rosenberg15,Mashian15}. Breaking this degeneracy requires spatially resolved line-ratio maps that can be directly compared with gas kinematics and the geometry of winds or jets.

In addition to the multiple CO transitions, a broader molecular-line census is also required to fully characterise the physical (e.g.\,density, temperature, ionisation state) and chemical gas conditions of the gas.  For instance, enhanced HCN/HCO$^+$ ratios are frequently observed in AGN-dominated regions and may reflect X-ray irradiation, abundance variations, or infrared radiative pumping \citep[e.g.][see, however, \citealt{Privon20}]{Imanishi14,Izumi16,Ruffa18}. However, shocks associated with molecular outflows can produce similar enhancements \citep[e.g.][]{Nishimura24,Imanishi26}, again demonstrating that individual line ratios cannot be interpreted independently of spatially resolved gas kinematics.

Atomic carbon ([C\,{\sc i}]) is another key diagnostic of gas affected by AGN feedback: in X-ray and cosmic-ray-dominated  environments, CO can be dissociated over large columns
while H$_2$ survives, leaving a CO-dark molecular component \citep{Papadopoulos04,Bisbas17}. In NGC~7469, for instance, ALMA reveals a strong nuclear enhancement of [C\,{\sc i}](1--0) relative to $^{13}$CO(2--1), attributed to X-ray irradiation and consequent heating \citep[][Fig.~\ref{fig:izumi20}]{Izumi20}, while in Circinus [C\,{\sc i}] traces an atomic outflow \citep{Zschaechner16}. Therefore, simultaneous observations of [C\,{\sc i}], CO and dense/shock gas tracers at matched resolution are needed to reduce systematic uncertainties on molecular gas masses, gas fractions and outflow rates.

Numerical models provide specific physical scenarios against which similar observations can be tested. Simulations of gravitationally-driven feeding predict a sequence of non-axisymmetric structures that can transport gas through the circumnuclear region, with the dominant mechanisms changing towards the SMBH \citep{Hopkins10}. On the other hand, simulations of jet--ISM interactions on (sub-)kpc scales predict gas cloud acceleration, compression, and enhanced turbulent motions, whose relative importance depends on jet power, relative jet--gas orientation, and the structure of the gas reservoir \citep{Mukherjee18a, Mukherjee18b}. 
In particular, the efficiency with which jets disperse the gas depends on the size of the largest cloud complexes, with structures of $\gtrsim$50\,pc requiring substantially more powerful jets \citep{Wagner12}. Hydrochemical models of AGN winds offer a complementary possibility: some of the observed molecular gas may form in cooling, swept-up material rather than being entirely entrained as pre-existing molecular clouds \citep{Richings18}. Testing these scenarios requires the joint distribution of gas kinematics and chemistry. Nearby benchmarks such as IC~5063, NGC~1068, Centaurus~A, and NGC~3100 have already demonstrated the power of this approach 
\citep[e.g.][]{Oosterloo17,Garcia19,Espada17,Espada19,Ruffa22}. ALMA2040 must extend this capability from individual local case studies to much larger samples, spanning broader ranges of distances and the full diversity of AGN classes. 

\subsection{The missing frontier: cold gas in kinetic-mode AGN}\label{sec:radio_AGN}
The application of the diagnostics outlined in Section~\ref{sec:fuel_feedback} is particularly urgent for kinetic-mode AGN hosted by massive ETGs. Up to at least $z\sim0.1-0.2$, the most massive galaxies are typically passive, early-types embedded in hot gaseous halos, where radio jets arising from low-accretion rate AGN deposit mechanical energy both on galactic and circumgalactic scales \citep[e.g.][]{McNamara12,Heckman14,Hardcastle20}. The latter case is known as ``maintenance-mode'' feedback, and it is typically invoked to prevent catastrophic cooling, suppress late-time star formation, and maintain the quiescence of massive galaxies \citep[e.g.][]{Sijacki07,Gaspari13,Weinberger17}. However, the cold molecular phase in these systems remains comparatively unexplored (see also Section~\ref{sec:ALMA_state}).

This is a critical gap. In kinetic-mode AGN, the same cold gas may act as fuel for the SMBH, as the target of jet-driven feedback on both (sub-)kpc and tens of kpc scales, and as the end product of thermally-unstable cooling from the hot atmosphere \citep{Gaspari13,Voit15,Tremblay16,Temi18,Mukherjee18b,Rose19,Ruffa19a,Ruffa22,Oosterloo24}. In brightest cluster galaxies and group-centred ETGs, molecular filaments on kpc or tens of kpc scales are often spatially associated with X-ray cavities, radio bubbles, and optical line-emitting nebulae, suggesting a close connection between jet activity, uplift, cooling, and subsequent accretion \citep{Russell14,Russell19,Tremblay18,Olivares23}. In lower-power radio galaxies hosted by ETGs, ALMA observations have revealed compact molecular disks and disturbed gas structures that may trace either externally accreted material, cooling from the hot halo, or jet-induced perturbations of a pre-existing gas reservoir on sub-kpc scales \citep[e.g.][see also \citealp{Ruffa24} for a review on this subject]{Ruffa19b,Ruffa19a,Ruffa20,Ruffa22, Rosario19, North21,Tadhunter24,Oosterloo25}.

Notably, where these circumnuclear disks are dynamically relaxed, they already serve double duty: the same ALMA cubes have yielded precise gas-dynamical SMBH masses in several nearby radio galaxies \citep[e.g.][]{Boizelle19, Ruffa19a, North19, Nguyen22, Ruffa23}, anchoring the jet powers and accretion rates of kinetic-mode AGN to directly measured black hole masses (see also White Paper of this series by Decarli et al.).

Despite these advances, we still lack a complete physical picture of how cold gas behaves across the kinetic-mode duty cycle.  The decisive advance with ALMA2040 will thus be to apply the method described in Section~\ref{sec:fuel_feedback} in sizeable samples of kinetic-mode AGN to probe the role of cold gas in their SMBH fuelling, fully determine the impact of mechanical AGN feedback in these systems and test how this depends on radio jet power, source age, and jet--disc relative orientation. Sensitive observations of molecular filaments in group- and cluster-centred kinetic-mode AGN will provide a complementary test of cooling from the hot atmosphere. 

\begin{tcolorbox}[
    colback=blue!5,
    colframe=blue!55!black,
    boxrule=0.8pt,
    arc=2pt,
    left=6pt,
    right=6pt,
    top=5pt,
    bottom=5pt
]
\small
\textbf{Closing the AGN molecular gas cycle: key questions for ALMA2040} 
\begin{itemize}
    \setlength{\itemsep}{2pt}
    \setlength{\parskip}{0pt}
    \setlength{\parsep}{0pt}
\item \textbf{The fate of the gas:} Does radiative/kinetic AGN feedback remove the molecular reservoir, or does most of the affected gas remain bound, stirred and eventually recycled? How does $\dot M_{\rm out}/\dot M_{\rm in}$ evolve over the AGN duty cycle?
\item \textbf{Self-regulation of feeding:} Does the AGN limit its own fuel supply by removing gas, or by preventing the angular-momentum loss required for inward
transport?
\item \textbf{The physical and chemical state of the gas:} How do radiative and/or mechanical energy injection from the AGN change the excitation, chemistry and CO-bright fraction of the molecular gas, and thus our estimates of gas masses and outflow rates?
\end{itemize}
\end{tcolorbox}

\vspace{-0.6em}

\section{ALMA2040: Technical Requirements}

\noindent Unveiling the AGN feeding and feedback cycle through multi-line diagnostics of the physics and kinematics of the cold ISM requires a next-generation sub-millimetre interferometer with a major leap in spectral-line sensitivity, mapping speed, and imaging fidelity.\,A factor of $\sim$10 improvement in instantaneous line sensitivity with respect to the current ALMA, corresponding to $\sim$8 relative to the anticipated ALMA-WSU, is essential to detect and accurately map faint molecular tracers such as HCN, HCO$^+$, CS, and SiO (typically ${\sim}10{-}50$x fainter than the $^{12}$CO lines), at different physical scales (from pc to kpc), and across statistically significant samples of AGN. Fast mosaicking capabilities are equally crucial to connect the nuclear regions, where feeding and feedback originate, with the circumnuclear and galaxy-wide cold ISM affected by AGN-driven energy injection. These requirements can be achieved primarily through a substantial increase in collecting area relative to both the current ALMA and the ALMA-WSU, together with disruptive developments in receiver technology. In particular, simultaneous multi-band observations and ultra-wide bandwidths would enable efficient, homogeneous and instantaneous coverage of multiple molecular transitions. Furthermore, the expected increase of the maximum baseline to $\sim$30\,km achievable within the Atacama astronomical park ($\sim2\times$ larger than the current ALMA) would yield synthesised beams of up to $\sim$9\,mas at 230\,GHz. This will provide a nominal physical resolution better than 50\,pc (i.e.\,the typical median size of single giant molecular clouds, GMCs) out to $z \simeq 0.4$, while genuinely resolving 50-pc-scale gas structures with multiple resolution elements at $z \lesssim 0.1$. Achieving spatial resolutions of $\sim 50$ pc is not strictly required to meet the primary science goals of the proposed studies (see, e.g., the case of NGC~3100 in Section~\ref{sec:ghost_sim}). However, such resolutions would be particularly valuable for disentangling spatially-overlapping gas components with distinct kinematics and excitation conditions, which may remain unresolved or blended at typical resolutions of a few hundred parsecs (see Section~\ref{sec:ghost_sim}).  At the same time, exquisite image fidelity, enabled by an improved $uv$ coverage (i.e.\,larger number of antennas) and better access to large angular scales, is required to recover both compact nuclear structures and diffuse extended emission (see Section~\ref{sec:radio_AGN}). 

\subsection{A view towards 2040: simulations with GHOST}
\label{sec:ghost_sim}

\begin{figure}[t!]
\centering
\includegraphics[clip=true, trim={2 2 2 2}, scale=0.54]{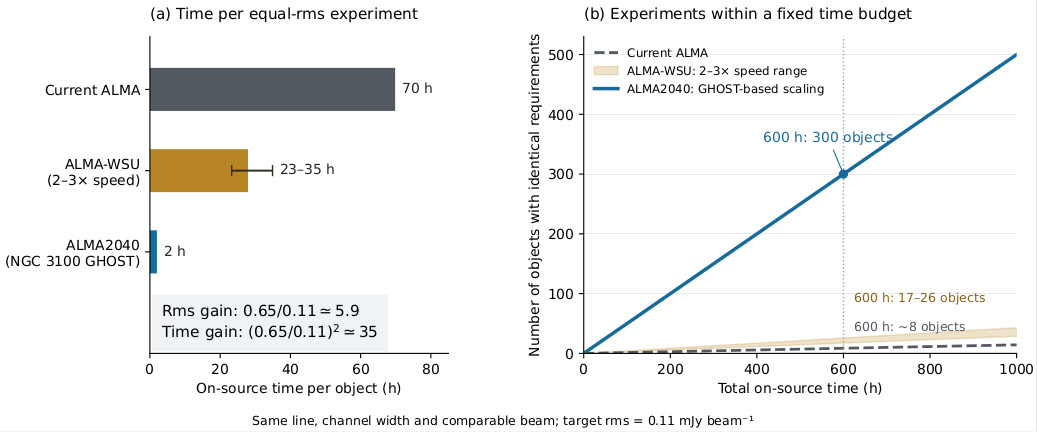}
\caption{\it (a) On-source time required to reach 0.11 mJy\,beam$^{-1}$ per channel at comparable beam and identical spectral sampling. The illustrative ALMA-WSU range applies the band-dependent factor of 2--3 single-line imaging-speed gain reported in \citet{2023pcsf.conf..304C}. (b) Number of equivalent observations as a function of total on-source time. All objects are assumed to have identical observational requirements. The reference experiment is the one described in Section~\ref{sec:ghost_sim}. The scaling excludes overheads and additional lines or configurations, and does not predict target availability.}\label{fig:census_scaling}
\end{figure}

To illustrate the transformative potential of ALMA2040, we used the ALMA GHOST simulator (\url{https://alma-ghost.jb.man.ac.uk}) to investigate how far the spatially-resolved molecular gas experiments described in this paper could be extended. We adopted the radio galaxy NGC\,3100 ($z=0.0088$) as a well-characterised benchmark in which multi-transition ALMA molecular gas observations at physical resolution of a few hundreds of pc have accurately established both the molecular-disc dynamics (including inflow/outflow motions) and changes in the gas excitation associated with a jet--ISM coupling on sub-kpc scales \citep{Ruffa19b,Ruffa19a,Ruffa20,Ruffa22}. Using the CO(3--2) kinematic model presented in \citet{Ruffa22}, we constructed a three-dimensional sky model, rescaled it to $z=0.05$, and simulated 2\,h observations with current ALMA and the ALMA2040 configurations available within GHOST, adopting identical observing conditions, spectral resolution and Briggs weighting ($R=0.5$). 
\begin{figure}[t!]
\centering
\includegraphics[scale=0.43]{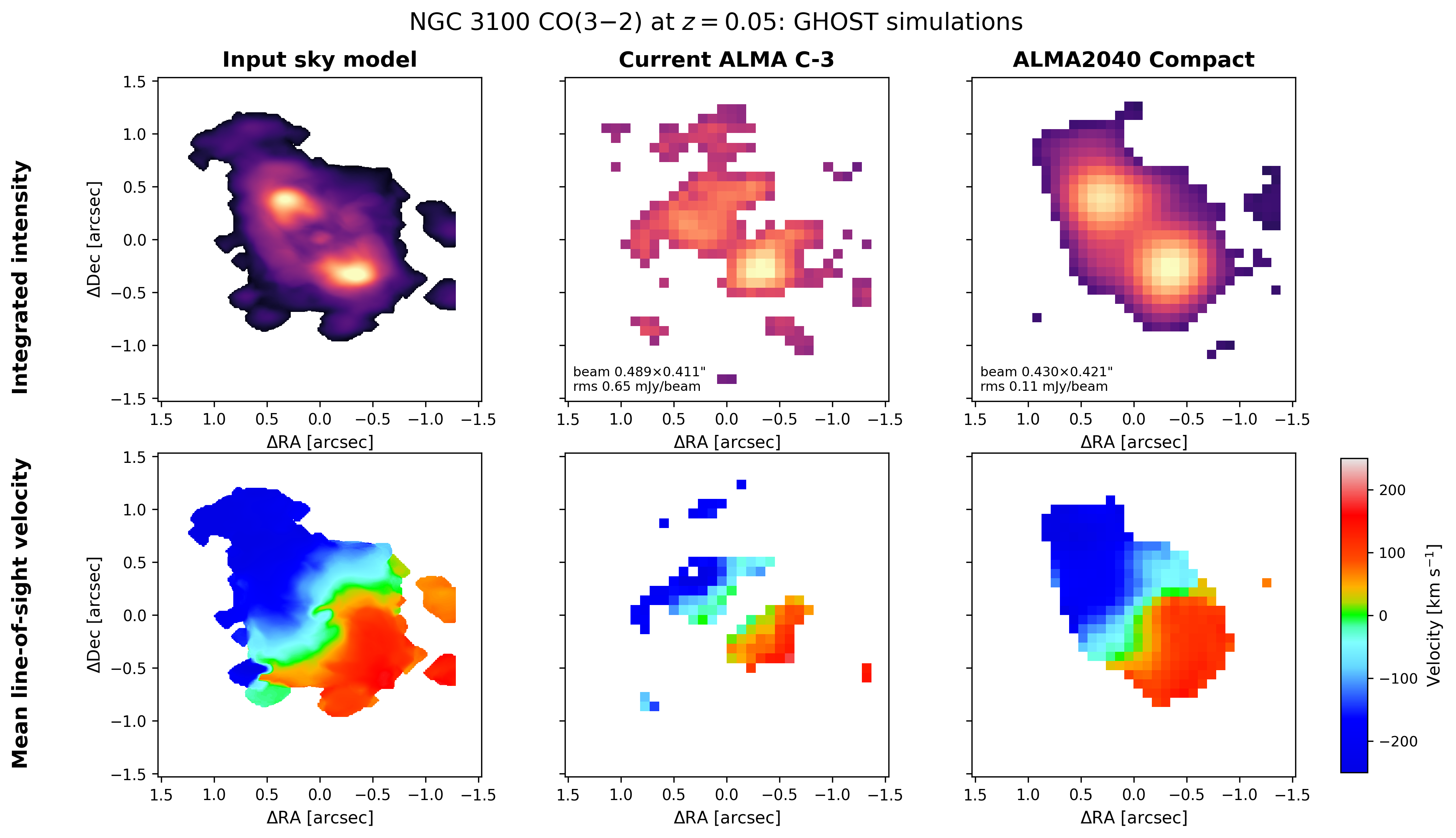}
\caption{\it GHOST simulations of an NGC~3100-like molecular disc. A realistic CO(3--2) model of the radio galaxy was rescaled to $z=0.05$ and observed for 2\,h under identical conditions with current ALMA C-3 and the proposed ALMA2040 Compact array. Top panels show the integrated emission maps and bottom panels the mean line-of-sight velocity fields.\,The left column shows the injected sky model. Current ALMA and ALMA2040 provide comparable angular resolutions ($0.49''\times0.41''$ and $0.43''\times0.42''$, respectively), while the channel rms improves from $\simeq0.65$ to $\simeq0.11$\,mJy\,beam$^{-1}$. The rms ratio implies an approximately 35-fold thermal-noise observing-speed gain at comparable angular resolution (see the text for details).}\label{fig:ghost_sim}
\end{figure}
Figure~\ref{fig:ghost_sim} compares the injected model with current ALMA C-3 configuration and the simulated ALMA2040 Compact array ($255\times12$-m antennas, $\sim0.4$\,km max baseline), providing comparable angular resolution ($0.49''\times0.41''$ and $0.43''\times0.42''$, respectively). The comparison prensented in Figure~\ref{fig:ghost_sim} clearly shows that, while the current array recovers only a fragmented representation of the velocity field, ALMA2040 reconstructs the full molecular disc and its resolved line-of-sight velocity structure with striking fidelity. The channel rms decreases from $\simeq0.65$ to $\simeq0.11$\,mJy\,beam$^{-1}$ in the same integration time. The rms ratio corresponds to a fixed-time line sensitivity gain of approximately 5.9 and, in the thermal-noise regime, an observing-speed gain of approximately 35. Current ALMA would therefore require approximately 70\,h on-source to reach the $1\sigma$ channel rms achieved by ALMA2040 in this 2\,h experiment. Thus, a campaign of 600\,h on-source time would enable observing around 300 such targets with ALMA2040, compared to only 8 with current ALMA (Fig.~\ref{fig:census_scaling}). The illustrative factor of 2--3 single-line imaging-speed improvement anticipated for ALMA-WSU would allow approximately 17--26 equivalent experiments in the same time budget \citep{2023pcsf.conf..304C}. These numbers clearly illustrate why ALMA2040 is needed to enable the transition from individual case studies to systematic comparisons within AGN classes. Similar studies require a potential sample to be divided into scientifically-defined bins: subsets of AGN sharing the properties relevant to the question being tested. For example, to investigate whether changes in the molecular-gas physics and kinematics are somehow correlated to the accretion mode and AGN luminosity, one could distinguish radiative- and kinetic-mode AGN and divide each population into three luminosity intervals, yielding six comparison bins. An illustrative sample of 300 objects would thus provide 50 AGN per bin, if these categories were equally populated. Within each bin, we could measure the fraction of systems showing a specified signature, such as a molecular outflow, disturbed kinematics or enhanced excitation, with a statistical uncertainty of at most $\approx7$\% ($1\sigma$). Comparing these fractions would establish whether a given response is common or exceptional and whether its incidence changes with different AGN properties/types. Note that the time scaling is illustrative: it applies to identical single-line requirements, excludes overheads, and is not a forecast for a complete multi-line census. The $\approx0.45''$ beam of the simulated ALMA2040 data corresponds to $\approx400$\,pc at $z=0.05$: the experiment presented here therefore demonstrates the recovery of disc-scale emission and kinematics, while finer resolution studies of NGC~3100-like cases at $z\gtrsim0.05$ would require additional simulations combining multiple ALMA2040 array configurations (an option that is currently not available in GHOST).

The simple example presented here clearly shows the tremendous potential of a major instrument upgrade such as ALMA2040: extending spatially-resolved molecular gas studies up to $z\simeq0.05$ would already dramatically increase the accessible volume and the number and diversity of AGN that can be studied at a high level of detail. Combining the sensitivity and angular resolution of the ALMA2040 12-m arrays with complementary compact-array and total-power information will also provide the required range of spatial scales to push such studies towards $z\sim0.1$ and beyond, while retaining practical observing times. ALMA2040 would therefore make it possible, for the first time, to build spatially-resolved molecular atlases of AGN host galaxies, probing different spatial scales and directly linking black-hole accretion, multiphase feedback, and the physical state of the cold ISM not merely for a few very nearby systems (i.e. up to $\sim100$\,Mpc), but for statistically significant samples of AGN spanning a wide range of nuclear activity levels, host morphologies and large-scale environments.

\section{Synergies with current and future facilities}
\noindent The study proposed in this paper will be a crucial step toward understanding how black-hole growth and galaxy evolution are coupled in the 2040s. By that time, the ELT will be in full operation, while JWST and large imaging and spectroscopic surveys from both the ground and space, including Euclid, Rubin, and Roman, will have provided an unprecedented legacy for tracing stellar mass assembly, ionised, and warm molecular gas. At the same time, new long-wavelength radio facilities such as the SKA, DSA-2000, and ngVLA will provide sensitive constraints on radio jets, non-thermal emission, and star formation. 
In this landscape, ALMA2040 will be uniquely positioned to deliver an unprecedented view of the cold ISM.

Among these facilities, the SKA will be the most direct complement to ALMA2040. Its $\mu$Jy-level sensitivity, sub-arcsecond resolution, and wide-field survey capabilities will enable large, homogeneous studies of H\,{\sc i} emission and absorption, tracing the neutral gas content, inflows, and outflows in AGN out to intermediate redshifts. Deep broadband continuum observations will also resolve radio jets over a wide range of scales, constraining their energetics and mechanical energy injection into the ISM \citep[see the SKA science cases of][]{Maccagni26, PandeyPommier2026}. Combined with ALMA2040 molecular-line diagnostics, these data will allow an unprecedented view of the molecular and atomic gas phases, and their correlation with the non-thermal radio jet component, providing a single multiphase picture of AGN feeding and feedback. 

The synergy with the ELT is also direct. HARMONI integral-field spectroscopy ($R\simeq7000$, spaxel scales down to $\sim$6\,mas with laser-guide-star adaptive optics) will deliver optical spectroscopy at a resolution matched to the ALMA2040 gas maps, providing in particular a complementary detailed view of the ionised-outflow and warm molecular gas phases in the central regions of AGN-host galaxies. 

\section*{Acknowledgements}
We thank the two referees, T.\,Diaz Santos and I.\,Prandoni, for their valuable comments that helped improving this paper. This work was developed as part of the broader ALMA2040 community effort. We acknowledge the Lorentz Center and NOVA for supporting the community workshops and coordination activities that helped initiate and shape the ALMA2040 White Paper process.\\
FS acknowledges financial support from the Bando Finanziamento ASI CI-UCO-DSR-2022-43 CUP: C93C25004260005 project ``IBISCO: feedback and obscuration in local AGN'', the MINI-GRANTS RSN1 "ECHOS", and the Ricerca Fondamentale INAF 2023 Data Analysis grant 1.05.23.03.04 ``ARCHIE ARchive Cosmic HI \& ISM  Evolution''.
AAH acknowledges support from grant PID2025-168017NB-C22.\\

\small
\bibliographystyle{apj_mod}
{\bibliography{bibliography}}
\end{document}